\documentclass[preprint2]{aastex}

\usepackage{amsmath,amssymb} 
\usepackage{float}

\shorttitle{Thermal tides}
\shortauthors{Goodman}

\begin{document}
\title{Concerning thermal tides on hot Jupiters}


\author{J. Goodman}
\affil{Princeton University Observatory, Princeton, NJ 08544}
\email{jeremy@astro.princeton.edu}

\begin{abstract}
  By analogy with a mechanism proposed by Gold and Soter to explain
  the retrograde rotation of Venus, Arras and Socrates suggest
  that thermal tides may excite hot jovian exoplanets into
  nonsynchronous rotation, and perhaps also noncircular orbits.  It is
  shown here that because of the absence of a solid surface above the
  convective core of a jovian planet, the coupling of the
  gravitational and thermal tides vanishes to zeroth order in the ratio
  of the atmospheric scale height to the planetary radius.  
  At the next order, the effect probably has the sign opposite to that
  claimed by the latter authors, hence reinforcing synchronous and circular orbits.
\end{abstract}




\goodbreak

\section{Introduction and Discussion}
\begin{figure}[H]
  \hspace*{-0.6in}
  \includegraphics[width=4in]{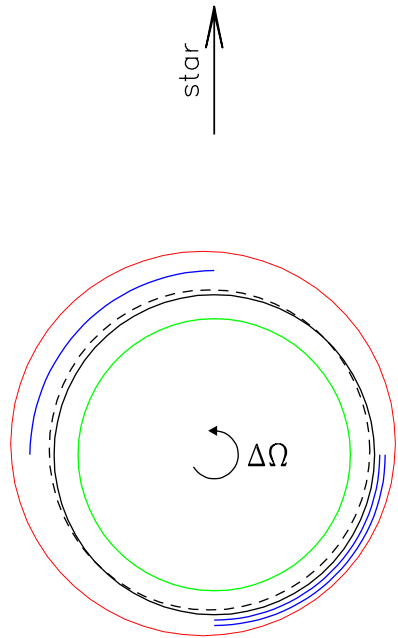}
  \caption{Thermal tide. Planetary
    rotation is nonsynchronous by amount $\Delta\Omega$.  Black solid
    circle is the surface (Venus) or the top of the undisturbed
    convective core (hot Jupiters).  Outermost (red) curve is an
    isobar: the atmosphere is hottest and most extended in the
    ``afternoon'' and coolest before dawn.  Lateral pressure gradients
    concentrate mass (blue arcs) toward colder regions.  Innermost
    (green) circle is an equipotential inside the convection zone
    (jovian case).  This zone being isentropic and hydrostatic,
    equipotentials coincide with isobars. Hence the mass column above
    the green circle is constant.  Redistribution of the atmosphere is
    isostatically compensated by shifts in the convective/radiative
    interface (dashed).  }
  \label{fig:fig1}
\end{figure}

The thermal tide in an orbitally circularized but
asynchronously rotating planet is sketched in Figure~\ref{fig:fig1}.
The figure emphasizes the diurnal component, that
is, the atmospheric variations that undergo one full period per
longitudinal circuit of the equator.  The gravitational tide couples
to the next Fourier harmonic, the semidiurnal tide.  This also reaches
a temperature maximum in the ``afternoon.'' To the extent that the
atmosphere rearranges itself into approximate
lateral pressure equilibrium, density variations are $180^\circ$ out of
phase with temperature variations.  For the semidiurnal component,
this puts the density maxima in the late morning and evening, i.e. at
$\pm 90^\circ$ of longitude with respect to the temperature
maxima.  The torque exerted on the thermal tidal density
by the gravitational tidal potential tends to accelerate the
asynchronous rotation of the atmosphere, as argued by
\cite{Arras_Socrates09}.

\cite{Gold_Soter69} proposed that the thermal tide in Venus'
atmosphere explains why the planet maintains a retrograde rotation
despite the fact that the gravitational tide acting on its solid parts
would be expected to lead to synchronism in $\sim 10^8{\rm yr}$.
Their theory was later refined by Ingersoll and Dobrovolskis
\citep{Ingersoll_Dobrovolskis78,Dobrovolskis_Ingersoll80}.  As these
papers make clear, the torque acting on the thermal tide is directly
proportional to the longitudinal pressure variation at the planetary
surface, which reflects the variation in atmospheric column density by
Pascal's Principle, since the atmosphere can be assumed to be in
vertical hydrostatic equilibrium to a good approximation.  Venus'
surface can withstand this longitudinal pressure variation because
it has elastic strength.  

A jovian planet, being gaseous, lacks elastic strength.  The
excess column density of the colder parts of the atmosphere is
counterbalanced---to the degree that hydrostatic equilibrium
holds---by an indentation of the convective boundary and a
redistribution of the core's mass toward the hotter longitudes.
Insofar as the radial range over which mass redistribution occurs is
small compared to the planetary radius, the thermal tide therefore
bears no net mass quadrupole.  The torque on the atmosphere is opposed
by a torque on the upper parts of the convection zone.

This compensation is analogous to geological isostasy; both are applications
of Archimedes' Principle \citep{Watts01}.  The earth's crust, or
lithosphere, floats upon the more plastic aesthenosphere, displacing
approximately its own weight.  Continental crust is less dense than
oceanic crust and floats higher, so that column density is
approximately constant.  To the extent that isostatic equilibrium
holds, therefore, gravitational anomalies experienced by a
satellite such as GRACE\footnote{{\tt
    http://www.csr.utexas.edu/grace/}} are of second order in crustal
density and thickness variations.  Whereas terrestrial isostasy operates
on such long timescales that rock behaves as fluid, the corresponding
timescale for gaseous planets is dynamical, hence less than the tidal
period.

Thus, the tidal torque claimed by \cite{Arras_Socrates09} vanishes to
first order in the density variations of the thermal tide.  To the
next order, the quadrupole moment of the thermal tide
aligns with the hottest and most distended parts of the atmosphere,
because mass elements are weighted by the squares of their distances
from the center.  This will lead to a torque of the
opposite sign to that of $\Delta\Omega$, hence driving the planet
toward synchronous rotation.  Similarly, the phase lag of the thermal
tide associated with an orbital eccentricity will affect the orbit
only to second order, and will tend to circularize the orbit.

Perfect lateral hydrostatic equilibrium cannot exist in atmospheres
where the temperature varies along isobars: winds must occur.
Furthermore, some mechanical friction or dissipation is required to
offset the tendency of the winds to accelerate \citep{Goodman08}. Thus
some differential rotation is expected, but not necessarily a
\emph{thermally} driven exchange of angular momentum between
rotation and orbit.

\vspace{10pt} 
This work was supported by the National Science Foundation
under grant AST-0707373.

\bibliographystyle{apj}
\bibliography{thermal}

\end{document}